\documentclass[preprint,showpacs,preprintnumbers,amsmath,amssymb,prl,aps]{revtex4}

\usepackage{epsfig}
\usepackage{graphicx}
\usepackage{dcolumn}
\usepackage{bm}

\begin{document}
\title{Role of long range ferromagnetic order in the electronic structure
of Sr$_{1-x}$Ca$_x$RuO$_3$}

\author{Ravi Shankar Singh, V.R.R. Medicherla, and Kalobaran Maiti\footnote{Author
to whom correspondence should be addressed; electronic mail:
kbmaiti@tifr.res.in}}

\affiliation{Department of Condensed Matter Physics and Materials
Science, Tata Institute of Fundamental Research, Homi Bhabha Road,
Colaba, Mumbai 400005, India}

\date{\today}

\begin{abstract}

We investigate the role of long range ferromagnetic order in the
electronic structure of Sr$_{1-x}$Ca$_x$RuO$_3$ using high
resolution photoemission spectroscopy. SrRuO$_3$ is a ferromagnetic
metal but isostructural, isoelectronic CaRuO$_3$ is an enhanced
paramagnet. Surface spectra of CaRuO$_3$ exhibit temperature induced
modifications. This is not significant in other compositions. This
may be attributed to the structural changes observed in previous
studies. Interestingly, the bulk spectra reveal unusual spectral
changes exhibiting large decrease in the coherent feature intensity
corresponding to only ferromagnetic samples, although the Ru moment
is very similar in all the compositions.

\end{abstract}


\maketitle

Ruthenium based perovskite oxides have attracted a great deal of
attention due to possibilities in significant technological
applications in addition to various interesting fundamental issues.
In particular, SrRuO$_3$, the only itinerant ferromagnet among the
4$d$ transition metal oxides (Curie Temperature, T$_c$ $\sim$ 165~K)
\cite{ref1}, is a promising candidate for several technological
applications due to its metallic character, high magnetic moment
(1.4 $\mu_B$/Ru), high chemical stability, etc. \cite{ref1,cao,DAE}
Ultrathin films of SrRuO$_3$ have already been used for normal metal
layers in Josephson junctions \cite{ref2}, spintronic devices based
on spin polarized ferromagnetic tunnel junctions with ferromagnetic
metal as an electrode \cite{ref3}, etc. Thus, ferromagnetic
materials form basis for technological advances and microscopic
understanding of the origin of such effect is crucial to design new
materials for future applications.

Here, we investigate the role of ferromagnetic transition in the
electronic structure of orthorhombically distorted perovskites
(space group $Pbnm$), Sr$_{1-x}$Ca$_x$RuO$_3$. The average Ru-O-Ru
bond angle gradually decreases from $\sim$~165$^\circ$ in SrRuO$_3$
to about 150$^\circ$ in CaRuO$_3$ \cite{ref1,DAE,kobayashi}.
Magnetic measurements exhibit ferromagnetic ground state for $x
<$~0.8 and enhanced paramagnetic phase for higher $x$ values
\cite{cao}. Such different magnetic ground states have also been
observed in {\em ab initio} calculations based on local spin density
approximations (LSDA) \cite{band2,band1}. Various photoemission
studies \cite{rapcom,ourepl,fujimori} suggest that the electron
correlation strength, $U/W$ ($U$ = Coulomb interaction strength, $W$
= bandwidth) is significantly weak and similar in all the
compositions. Transport measurements, on the other hand, indicate
Fermi liquid behavior in SrRuO$_3$ while CaRuO$_3$ is non-Fermi
liquid \cite{nfl}. It is thus, clear that these systems exhibit
varieties of interesting ground state properties, which cannot be
attributed solely to the electron correlation effect. Despite
numerous studies, the origin of such widely different ground state
properties is still unclear.

In this letter, we report our results on the modification of the
electronic structure across the magnetic phase transition in this
system using state of the art high resolution photoemission
spectroscopy. Experimental spectra exhibit qualitatively different
bulk and surface electronic structures in all the samples and
interesting evolutions with the change in temperature.

Samples were prepared by solid state reaction route followed by
sintering in the pellet form for about 72 hours at 1523~K to achieve
large grain size. Sharp features in the $x$-ray diffraction (XRD)
patterns with lattice parameters similar to those in single
crystalline samples \cite{cao,kobayashi} and no signature of
impurity feature indicate high quality of the samples. DC magnetic
susceptibility, measured using high sensitivity vibrating sample
magnetometer, show sharp ferromagnetic transition at 165~K in
SrRuO$_3$. The sharpness of the transition gradually reduces and
becomes insignificant for $x \geq$~0.8. $\mu$$_{eff}$ for all the
samples in the paramagnetic region has been estimated to be
2.8~$\pm$~0.2 $\mu$$_{B}$, which is close to the theoretical spin
only value of 2.83~$\mu_B$ corresponding to
t$_{2g\uparrow}^3t_{2g\downarrow}^1$ configuration of Ru$^{4+}$
\cite{cao,rapcom}. Photoemission measurements were performed on {\em
in situ} (base pressure~$\sim$~ 3$\times10^{-11}$ torr) scraped
sample surfaces using Gammadata Scienta analyzer, SES2002 with an
energy resolution set to 4~meV, 900~meV and 300~meV for the
measurements with monochromatic He {\scriptsize II} (40.8~eV), Al
$K\alpha$ (1486.6~eV) (twin source) and Al $K\alpha$ (monochromatic
source) respectively. Cleanliness of the sample surface was ensured
by minimizing the higher binding energy feature in O 1$s$ spectra
\cite {solidstate} and the absence of C 1$s$ signal. A
polycrystalline silver was mounted on the same sample holder in
electrical contact with other samples to determine the Fermi edge,
$\epsilon_F$.


In Fig.~1(a), we plot Ru 4$d$ He {\scriptsize II} spectra after
subtracting the O 2$p$ contributions appearing at higher binding
energies. All the spectra exhibit an intense, broad feature at
1.2~eV along with finite intensity at $\epsilon_F$. While the
intensities at $\epsilon_F$ correspond well to the band structure
results (termed as {\em coherent feature}), signature of the
dominant contributions at 1.2~eV is not present in these results
\cite{band2,band1}. The 300~K spectra for all the values of $x$ are
very similar exhibiting weak coherent feature intensity suggesting
metallic phase in these materials. Ru 4$d$ He {\scriptsize II}
spectra at 20~K exhibit significant decrease in intensity of the
coherent peak, when it is normalized by the intensity at 1.2~eV of
300 K spectrum. The difference in intensity between the spectra at
room temperature and 20~K spectra gradually increases with the
increase in $x$.

Since, the surface contribution is significantly large ($\sim$80\%)
in the He {\scriptsize II} spectra, we probe the influence of the
temperature on the Ru 4$d$ contributions in the Al $K\alpha$ spectra
of the valence band, where the surface sensitivity is reduced to
about $\sim$ 40\%. All the spectra are shown in Fig.~1(b) after
normalizing by the intensity at 1.5~eV. The lineshape of the Ru 4$d$
spectra is significantly different from the He {\scriptsize II}
spectra shown in Fig.~1(a). The 300~K spectrum of SrRuO$_3$ exhibits
intense coherent peak around 0.5~eV and an asymmetry towards higher
binding energies. Interestingly, corresponding 20~K spectrum
exhibits significant lowering in intensity compared to the intensity
at 1.5~eV. The difference in coherent feature intensity is
significantly large in SrRuO$_3$ and becomes almost insignificant in
CaRuO$_3$. This temperature induced modification is strikingly
different from that observed in the He {\scriptsize II} spectra.
This is verified in the high resolution spectra of end members shown
in Fig.~1(b). The high resolution spectra of SrRuO$_3$ exhibits a
large decrease in the coherent feature intensity with the decrease
in temperature, while CaRuO$_3$ spectra remain unchanged.


In order to understand the contrasting spectral changes in the He
{\scriptsize II} and Al $K\alpha$ spectra, we extract the surface
and bulk spectral functions from these two sets of spectra at all
the temperatures for all the samples. Photoemission intensity can be
expressed as $I(\epsilon) = [1 - e^{-d/\lambda}]F^s(\epsilon) +
e^{-d/\lambda}.F^b(\epsilon)$, where $d$ is the thickness of the
surface layer and $\lambda$ is the escape depth of the
photoelectrons. $F^s(\epsilon)$ and $F^b(\epsilon)$ represent the
surface and bulk spectra, respectively. Using the values of
$d/\lambda$ from Ref. [8], we have extracted the surface and the
bulk contributions as shown in Fig.~2(a) and Fig.~2(b),
respectively. The surface spectra exhibit dominant contributions at
1.2~eV binding energy. The threefold degeneracy of the Ru 4$d$
$t_{2g}$ band is already lifted in the bulk electronic structure due
to the distortion of the RuO$_6$ octahedra \cite{band1}. The absence
of periodicity along the surface normal will further reduce the
crystal symmetry from $O_h$ symmetry towards $D_{4h}$ symmetry at
the surface. Thus, the feature at 1.2~eV is often attributed to the
$e_g$ band derived from the $t_{2g}$ band due to such symmetry
breaking \cite{rapcom,surfbulk}.

The coherent feature intensity is significantly weak in the surface
spectra of all the compositions. The decrease in temperature down to
20~K does not lead to significant change in the surface spectra of
SrRuO$_3$ and Sr$_{0.7}$Ca$_{0.3}$RuO$_3$. This is also evident in
the high resolution spectra of SrRuO$_3$. Small change in lineshape
is observed for higher $x$ values, which is most significant in
CaRuO$_3$ exhibiting a large reduction in coherent feature intensity
with the decrease in temperature as clearly visible in the high
resolution spectra of CaRuO$_3$. Various core level studies
\cite{solidstate,Ca2pSr3d} indicate significant change in the
lineshape suggesting temperature induced modification in structural
parameters. It is already clear that the two dimensional nature,
defects, reconstructions at surface play key roles in determining
the surface electronic structure. Thus, the spectral modifications
observed in the surface spectra may be attributed to such
temperature induced changes of the surface structure.

In Fig.~2(b), we show the bulk spectra for all the $x$ values. The
room temperature spectra, normalized by integrated intensity under
the curve are almost similar in all the compositions. The coherent
peak appears at about 0.5~eV with the contribution of incoherent
peak (the lower Hubbard band) appearing at 2~eV. This suggests that
the change in Ru-O-Ru bond angle across the series does not
introduce significant change in the electronic structure of this
system. Bulk spectra at 20~K are shown in the same figure by
normalizing the intensity of the incoherent feature. Intensity of
the coherent feature in SrRuO$_3$ is found to decrease significantly
with the decrease in temperature across the magnetic phase
transition. Interestingly, such lowering in coherent features
intensity is clearly visible in the bulk spectra of all the
compositions exhibiting long range ferromagnetic order. The spectra
corresponding to $x \geq$~0.8 remain unchanged down to the lowest
temperature studied.

Band structure calculations \cite{band1} for various magnetic and
non-magnetic solutions suggest that in the ferromagnetic ground
state, the contribution from the down spin density of states moves
above the Fermi level due to the exchange coupling between the 4$d$
electronic states. Since, the coherent feature represents the
density of states observed in the band structure results, the
lowering of coherent feature intensity across the magnetic phase
transition may be attributed to the shift of down spin spectral
intensity above $\epsilon_F$. This shift of the down spin density of
states depends on the exchange splitting, which is also reflected in
the magnetic moment. This appears to explain the change in the
electronic structure in ferromagnetic compositions. However, various
magnetic measurements suggest similar Ru 4$d$ moment across the
entire series \cite{cao,rapcom}. Thus, no change in the bulk spectra
of paramagnetic samples is curious and opens up an interesting
question in microscopic understanding of the evolution of magnetism.

The ferromagnetism is often described within two models. (a) The
Stoner description \cite{stoner}: the exchange splitting gradually
decreases with the increase in temperature and becomes zero at the
Curie temperature leading to zero magnetic moment. (b) On the other
hand, a spin mixing behavior \cite{spinmix} leads to a reduction in
spin polarization with the increase in temperature keeping the
magnetic moment unchanged. The purely Stoner behavior can be ruled
out since the magnetic moment exists even in the paramagnetic phase
in all the compositions. If the second case is active, the spin
integrated spectra should be identical in all the compositions.
Thus, significant change only in the spectra of ferromagnetic
compositions in this study is curious and opens up a {\em new
dimension} in understanding ferromagnetism. Since, all the compounds
are essentially identical except the difference in long range order,
the significant modification observed in the samples having long
range order naturally suggests a relation among themselves. This
suggests that in addition to the intrasite exchange interactions
(responsible for local magnetic moments), intersite exchange
correlations, which give rise to long range order presumably play a
key role in determining the spectral functions observed by
photoemission spectroscopy. We hope, this study will help to
initiate further efforts in this direction to understand this effect
in ferromagnetic materials.

In summary, we investigate the change in the electronic structure
across the magnetic phase transition in a series of compounds
exhibiting magnetic ground states ranging from ferromagnetic to
enhanced paramagnetic. Although the intrasite exchange interactions
are similar in all the compositions, the bulk spectra exhibit
significant modification in the lineshape across the Curie
temperature in ferromagnetic materials, while the spectra in
paramagnetic samples remain unchanged down to the lowest temperature
studied. This suggests that intersite exchange interactions
responsible for long range order presumably play an important role
in determining the electronic structure of these interesting
materials.

\pagebreak

\pagebreak

\section{Figure Captions}

Fig. 1: Photoemission spectra of Ru 4$d$ valence band for different
values of $x$ in Sr$_{1-x}$Ca$_x$RuO$_3$ at 300~K (closed circle)
and 20~K (open circle) using (a) He {\scriptsize II} and (b) Al
$K\alpha$ radiations. The top and bottom sets in (b) are the high
resolution spectra of SrRuO$_3$ and CaRuO$_3$ using monochromatic Al
$K\alpha$ source.

Fig. 2: (a) Surface and (b) Bulk spectra of
Sr$_{1-x}$Ca$_{x}$RuO$_{3}$ at 300~K (closed circle) and 20~K (open
circle). Top and bottom sets are the high resolution spectra of
SrRuO$_3$ and CaRuO$_3$, respectively.


\begin{thebibliography}{99}

%
\bibitem{ref1} J. J. Randall and R. Ward, J. Amer. Chem. Soc. {\bf 81},
2629 (1959); A. Callaghan, C. W. Moeller, and R. Ward, Inorg. Chem.
{\bf 5}, 1572 (1966); J. M. Longo, P. M. Raccah, and J. B.
Goodenough, J. Appl. Phys. {\bf 39}, 1327 (1968).
%
\bibitem{cao} G. Cao, S. McCall, M. Shepard, J. E. Crow, and R. P. Guertin,
Phys. Rev. B {\bf 56}, 321 (1997).
%
\bibitem{DAE} R. S. Singh, P. L. Paulose, and K. Maiti, Solid State
Physics (India) {\bf 49}, 876 (2004).
%
\bibitem{ref2} S. C. Gausepohl, M. Lee, L. Antognazza, and K. Char,
Appl. Phys. Lett. {\bf 67}, 1313 (1995).
%
\bibitem{ref3} K. S. Takahashi, A. Sawa, Y. Ishii, H. Akoh, M. Kawasaki, and Y.
Tokura, Phys. Rev. B {\bf 67}, 094413 (2003).
%
\bibitem{kobayashi} Kobayashi, H., M. Nagata, R. Kanno, and Y. Kawamoto, Mater.
Res. Bull. {\bf 29}, 1271 (1994).
%
\bibitem{band2} D.J. Singh, J. Appl. Phys. {\bf 78}, 4818 (1996);
I.I. Mazin and D.J. Singh, Phys. Rev. B {\bf 56}, 2556 (1997).
%
\bibitem{band1} K. Maiti, Phys. Rev. B {\bf 73}, 235110 (2006).
%
\bibitem{rapcom} K. Maiti and R. S. Singh, Phys. Rev. B {\bf 71}, 161102(R)
(2005). In this paper, CaRuO$_3$ was described to be
antiferromagnetic due to (-ve) $\theta_P$ derived from the
susceptibility in the paramagnetic region. However, recent studies
suggest an enhanced paramagnetic phase in this compound.
%
\bibitem{ourepl} K. Maiti, R. S. Singh, and V. R. R. Medicherla,
Europhys. Lett. {\bf 78}, 17002 (2007).
%
\bibitem{fujimori} M. Takizawa, D. Toyota, H. Wadati, A. Chikamatsu,
H. Kumigashira, A. Fujimori, M. Oshima, Z. Fang, M. Lippmaa, M.
Kawasaki, and H. Koinuma, Phys. Rev. B {\bf 72}, 060404(B) (2005)
%
\bibitem{nfl} L. Klein, L. Antognazza, T. H. Geballe, M. R. Beasley
and A. Kapitulnik, Phys. Rev. B {\bf 60}, 1448 (1999); P. Khalifah,
I. Ohkubo, H. Christen and D. Mandrus, Phys. Rev. B {\bf 70}, 134426
(2004); Y. S. Lee et al., Phys. Rev. B {\bf 66}, 041104(R) (2002).
%
\bibitem{solidstate} R. S. Singh and K. Maiti, Solid State Comm. {\bf 140}, 188 (2006).
%
\bibitem{surfbulk} K. Maiti, A. Kumar, D. D. Sarma, E. Weschke, and G. Kaindl,
Phys. Rev. B {\bf 70}, 195112 (2004).
%
\bibitem{Ca2pSr3d} R. S. Singh and K. Maiti, cond-mat/0605552
%
\bibitem{stoner} E.C. Stoner, Proc. R. Soc. London A {\bf 154}, 656 (1936).
%
\bibitem{spinmix} V. Korenman {\em et al.}, Phys. Rev. B {\bf 16}, 4032 (1977);
{\bf 16}, 4048 (1977); H. Capellman, Z. Phys. B {\bf 34}, 29 (1979);
A.J. Pindor {\em et al.}, J. Phys. F {\bf 13}, 979 (1983); H.
Hasegawa, J. Phys. Soc. Jpn. {\bf 46}, 1504 (1979).
%

\end{thebibliography}
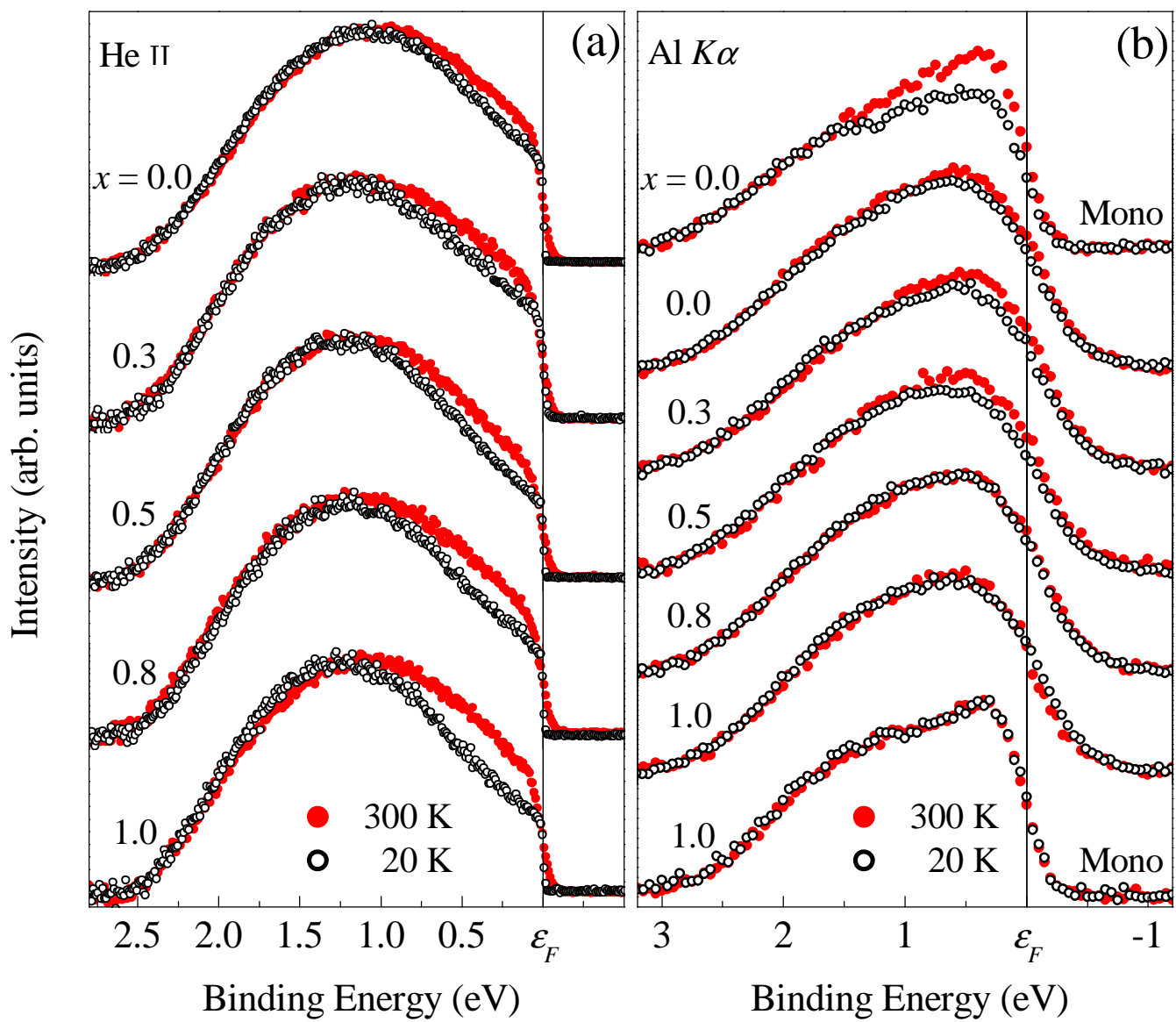
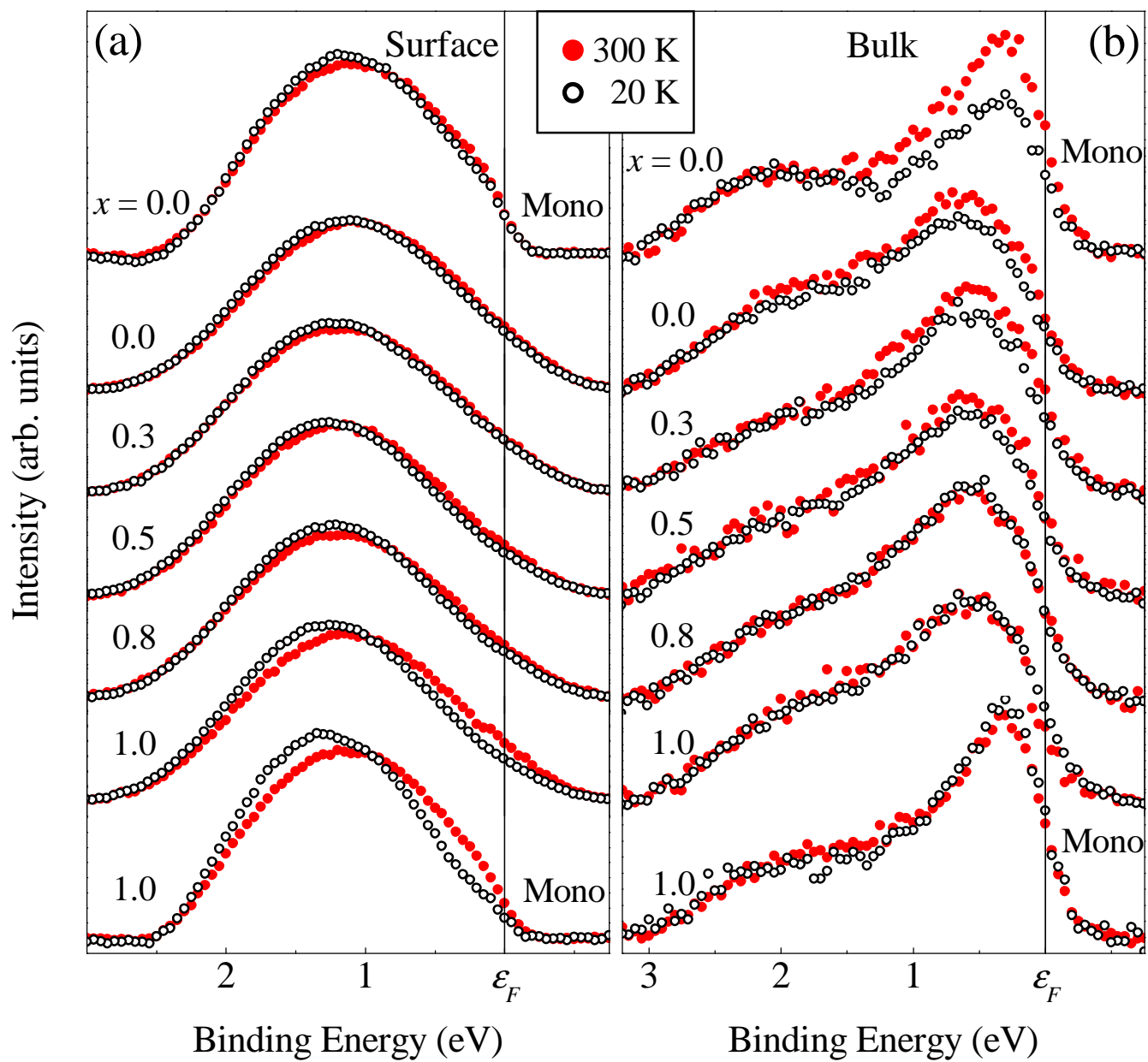
\end{document}